# A FURTHER SEARCH FOR GALACTIC STARS WITH DOUBLE RADIO LOBES

Braulio Arredondo Padilla[1] and Heinz Andernach[2]

## RESUMEN

Se sabe de más de mil estrellas en nuestra galaxia que emiten en ondas de radio, pero no se han encontrado estrellas normales que posean radiolóbulos similares a los de las radiogalaxias. Intentos recientes nuestros y de otros autores de encontrar tales objetos no han sido concluyentes. Aquí presentamos una nueva búsqueda de radioestrellas de doble lóbulo en dos muestras recientes de estrellas espectroscópicas: más de 20,000 enanas blancas del *Sloan Digital Sky Survey* (SDSS) DR12, y 2,540,000 estrellas del *Large Sky Area Multi-Object Fiber Spectroscopic Telescope* (LAMOST). Una búsqueda de pares de fuentes del rastreo *Faint Images of the Radio Sky at Twenty Centimeters* (FIRST) a 1.4 GHz con una simetría moderada respecto a dichas estrellas, resultó en sólo 4 candidatos, confirmando que deben ser extremadamente raras. Una comparación con el SDSS reveló que un 16 porciento de los espectros LAMOST pueden ser afectados por clasificaciones erróneas. Redescubrimos la radiogalaxia gigante J0927+3510 y proponemos aquí un huésped diferente, más distante, sugiriendo que su tamaño en radio alcanza los 2.7 Mpc.

## ABSTRACT

Over a thousand stars in our Galaxy have been detected as radio emitters, but no normal stars are known to possess radio-emitting lobes similar to radio galaxies. Several recent attempts by us and other authors to find such objects remained inconclusive. Here we present a further search for double-lobed radio stars in two large samples of spectroscopic stars: over 20,000 white dwarves from the *Sloan Digital Sky Survey* (SDSS) DR12, and 2.5 million stars from the *Large Sky Area Multi-Object Fiber Spectroscopic Telescope* (LAMOST). These were cross-matched with sources from the *Faint Images of the Radio Sky at Twenty Centimeters* (FIRST) survey at 1.4 GHz to look for source pairs straddling the stars with moderate symmetry about the stars. We found only four promising candidates for double-lobed radio stars, confirming they must be extremely rare. By comparison with SDSS, we inferred that about 16 per cent of LAMOST spectra may have erroneous classifications. We also rediscovered the giant radio galaxy J0927+3510 and propose a different, more distant host, suggesting a much larger radio size of 2.7 Mpc.

## INTRODUCTION

The first "discrete" radio sources found in the late 1940s were assumed to be "radio stars". By the early 1960s it was clear that virtually all are extragalactic. Apart from the detection of the Sun as a radio emitter, independently by Reber and Hey in 1944, it was not until 1970 that Galactic stars were detected in radio (Hjellming 1988). The latest compilation of radio-emitting stars (Wendker 2001) lists 1128 detected stars, half of which are located within 15° from the Galactic plane, and only 197 lie beyond 30° from it. An inspection by one of us (HA) of the ~120 of the latter stars that are covered in the FIRST survey revealed ~25 point sources, but no radio double at all.

The radio emission of all these stars must be due to rather violent processes, since the flux of the quiet Sun is well below the detection limit of even the next-generation radio telescopes. Moreover, the radio structure of the vast majority of these stars is unresolved, except for Herbig-Haro (HH) objects around newly born stars and the so-called microquasars (MQs, Mirabel and Rodríguez 1999), so far the only known stars with double lobes. Most HH host stars are not seen in the optical while MQs all have dark companions (black holes or neutron stars) causing an ejection of radio lobes via mass donation from the visible star to the dark companion, creating an accretion disk around the latter. Both, HHs and MQs, are close to the Galactic plane, and MQs are also copious X-ray emitters.

The search for further radio-emitting stars was motivated by the various data releases of the 1.4-GHz FIRST radio survey (1995−2015) eventually covering one quarter of the sky at an angular resolution of 5.4" (Helfand, White & Becker 2015). Helfand et al. (1999) searched for radio point sources in either bright or nearby stars, not finding any radio double sources. Kimball et al. (2009) searched for radio emission of fainter stars (15 < i < 19.1), only finding 112 unresolved sources, a number compatible with chance coincidence. These authors also reported another 76 stars with complex radio emission, but required the presence of a FIRST radio source within 1" from the star, thus excluding the possibility of the presence of double radio lobes with an extinct radio nucleus coincident with the star. Soon after the launch (in late 2013) of the citizen science project "Radio Galaxy Zoo" (RGZ, radio.galaxyzoo.org) aimed at the identification of complex FIRST radio sources on images of the WISE mid-infrared survey (later followed by AllWISE; Cutri et al. 2013), its volunteers occasionally reported double radio sources straddling bright stars. One striking example was FBQS J170008.6+291904, the core of a 2.63' wide radio triple, identified by White et al. (2000) as a r'=16.5$^m$ spectroscopic star. However, the active galaxy nucleus (AGN) reveals itself as an AllWISE source with non-stellar colors, typical for an AGN: W12,W23,W34=+0.58,+2.21,+1.5, yet unreported in literature. Moreover, the AllWISE position is ~0.2" from the FIRST position, while the star is 0.5" away from the FIRST position, suggesting that a background AGN within 0.5" of the star might be the real host.

[1] Universidad de Guanajuato, Departamento de Estudios Organizacionales, DCEA; Fraccionamiento 1 s/n, Col. El Establo, C.P: 36250, Guanajuato, Gto., b.arredondopadilla@ugto.mx
[2] Universidad de Guanajuato, Departamento de Astronomía, DCNE; Callejón. de Jalisco s/n, Col. Valenciana, C.P: 36240, Guanajuato, Gto., heinz@astro.ugto.mx

One of us (Andernach 2015) cross-matched the SAO and UCAC4 star catalogs to search for FIRST radio sources straddling the stars within 30" distance and with moderate symmetry, resulting in five stars brighter than V=11 mag. The suspected radio lobes are often slightly extended along the axis connecting the two radio sources, suggesting an ejection from the star. Unfortunately these stars are too bright to allow an AGN to outshine the stellar emission in the WISE mid-IR bands as for FBQS J170008.6+291904 above. All five objects have steep radio spectra and some have linear polarization detected, both rather typical features for distant radio galaxies. Moreover, the location of the stars near the axis of double radio sources of that size is entirely compatible with chance coincidence. Later Jimenez Valencia and Andernach (2015) searched for such double-lobed sources around stellar objects in the 2MASS Point Sources Catalog finding no convincing candidate. Ortiz Martínez and Andernach (2016) searched for radio doubles around 878,031 spectroscopic stars from SDSS DR12, but among their best three candidates, only a single one (SDSS J132941.25+151021.8) has AllWISE colors compatible with a star. The vast majority of candidate alignments of stars and double radio sources can be explained by faint galaxy hosts seen in SDSS, very close to the stars. Moreover, none of the potential double-lobed radio stars found have been detected in X-rays, all lie at high Galactic latitude and most of them are of spectroscopic type F or later, properties all very different from microquasars.

Here we extend this search to two large samples of spectroscopic stars, from Anguiano et al. (2017) and LAMOST DR2+DR3 (http://dr3.lamost.org). While SDSS generally only takes spectra of stars fainter than ~18$^{th}$ mag, LAMOST targets stars brighter than this, nicely complementing our previous searches for double-lobed radio stars.

## METHODS AND MATERIALS

Our search for radio doubles aligned with stars made use of a FORTRAN code written by M. A. Jimenez Valencia (Univ. de Sonora, México) during a summer research project with one of us (HA) in 2015. The code (MJV in what follows) calculates for all pairs of radio sources in a given field: (a) the angular distances from the star, (b) the orientation (position angle PA) of the vector from the central object to the radio source (c) the "armlength ratio" (ALR) as the ratio of stronger-to-weaker lobe distance from the star, (d) the difference between the two position angles (the "misalignment" or bending angle, or DPA), and (e) the ratio of their integrated flux densities as listed in the FIRST catalog, i.e. the flux ratio, FLR, in the sense of longer-to-shorter lobe flux. In the present work we concentrated on the most symmetric and aligned objects, with ALR and FLR closest to unity and DPA closest to 0°.

For our search we selected the compilation of 20247 confirmed white dwarf (WD) stars by Anguiano et al. (2015), as well as the 2.95 million spectra of spectroscopic stars contained in DR2 and DR3 of LAMOST (Cui et al. 2012). Using TOPCAT (Taylor 2013) we reduced the latter number to 2.54 million, by keeping only one of all duplicates within 2.5", assuming these to be repeat spectra. We then extracted FIRST radio sources around these stars, and considered only those stars with two or more matches within a radius of 60".

## RESULTS

### Serendipitous rediscovery of a giant radio galaxy

The WD catalog by Anguiano et al. (2015) has 151 WDs with two or more FIRST sources within 60". Applying the MJV code to these sources only a single WD (SDSS J092738.01+351126.4) survived our criteria for modest source symmetry with respect to the star position. This source pair consists of a compact FIRST source 13.5"W and a diffuse one 18"E of the WD. As the latter is listed in the FIRST catalog with a 28% probability of being a sidelobe, we inspected the surroundings of the pair, which turned out to be real, and in fact being the NW lobe of the giant radio galaxy (GRG) J0927+3510 as reported by Machalski et al. (2001) who proposed the r'=21.12 galaxy SDSS J092750.59+351050.7, undetected in AllWISE, as the GRG host. However, in our search for the most likely host we found SDSS J092749.75+351051.7, r'=23.09, with AllWISE colors of W12,W23=+0.27,+3.4. The previously proposed host has PanSTARRS PSFmags of gri=22.23, 21.86,21.55 (Flewelling et al. 2016), and is undetected in z and y, while our proposed host has rizy=21.94, 21.61, 20.91, 20.67. We retrieved the 8.46-GHz VLA radio image of Machalski et al. (2006) from the VLA archive and found that our proposed host coincides with a 0.17-mJy radio core, at least 5 times the rms noise in the image (0.03 mJy/beam), while the host proposed earlier is undetected at 8.46 GHz. The radio core position quoted in their table 2 does not coincide with either of the two optical objects and is based on their 4.86-GHz image with only 17" resolution. Moreover, our proposed host (large cross in Fig. 1) falls decidedly closer to the geometric center of the two inner hotspots than that proposed earlier (small cross E of center). The estimated redshift of the previous host candidate is 0.55, while that of the one proposed here is 0.937 (Abolfathi et al. 2017). If the latter is correct, the projected radio size (5.8') of this giant is closer to 2.7 rather than 2.2 Mpc ($H_0 = 70$ km s$^{-1}$ Mpc$^{-1}$, $\Omega_m=0.3$, $\Omega_\Lambda=0.7$). All images in this paper have north up and east to the left.

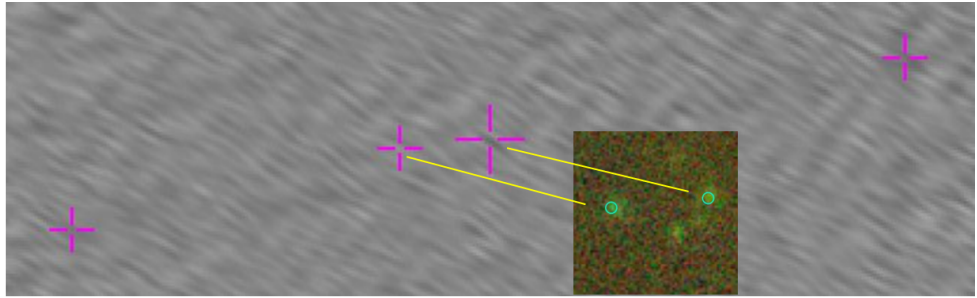

*Figure 1. 8.46-GHz VLA image (112" x 34") of the inner lobes of GRG J0927+3510. The large cross at center marks our proposed host, the smaller cross left of it is that proposed by Machalski et al. (2006). Red crosses at far left and right are peak positions of the E and W inner lobes of the GRG. The inset shows a 20" x 20" yig color composite from PanSTARRS with the two proposed hosts marked with green circles.*

**Potential Double-lobed Radio Stars in LAMOST DR2+3**

Of the 2.54 million spectroscopic stars in LAMOST DR2+3, 9052 stars have two or more (a total of 20410) FIRST sources within a radius of 60". Applying the MJV code to these we found 229 source pairs with reasonable source symmetry with respect to the stars, but we limited our visual inspection to the 78 most promising candidates with DPA < 5°, armlength ratio between 0.5 and 2.0, and lobe flux ratio between 0.2 and 5. After inspection of SDSS images and AllWISE colors, our best candidates are shown in Figure 2. Any extragalactic host for these must be either fainter than the PanSTARRS detection limit, or located within ~1" from the star and outshone by it.

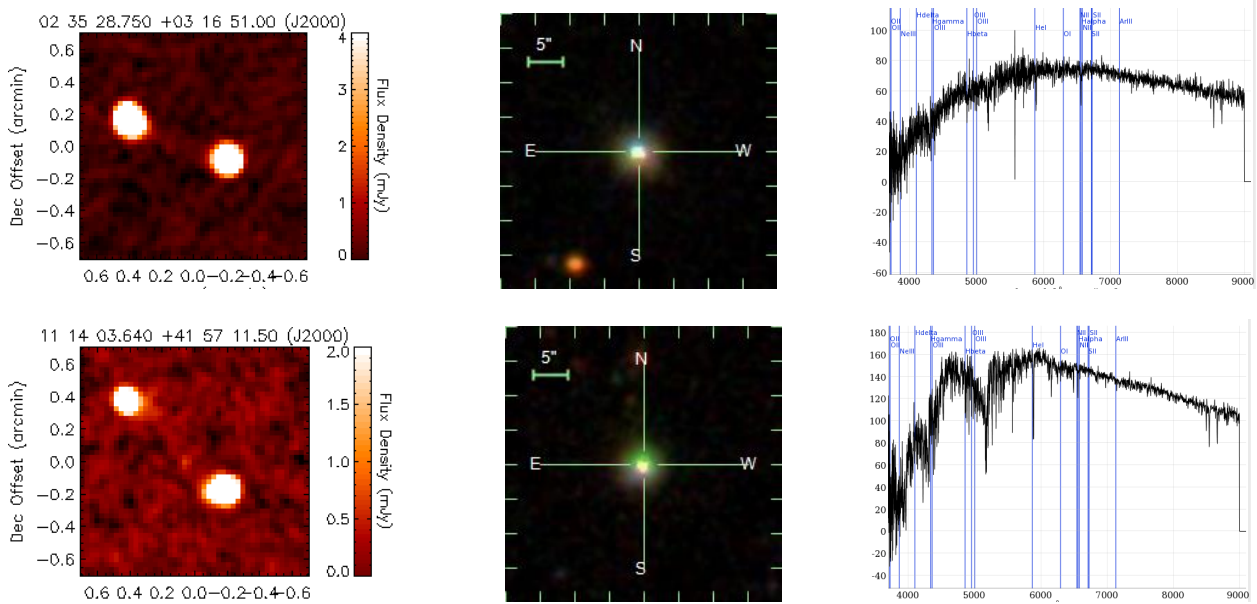

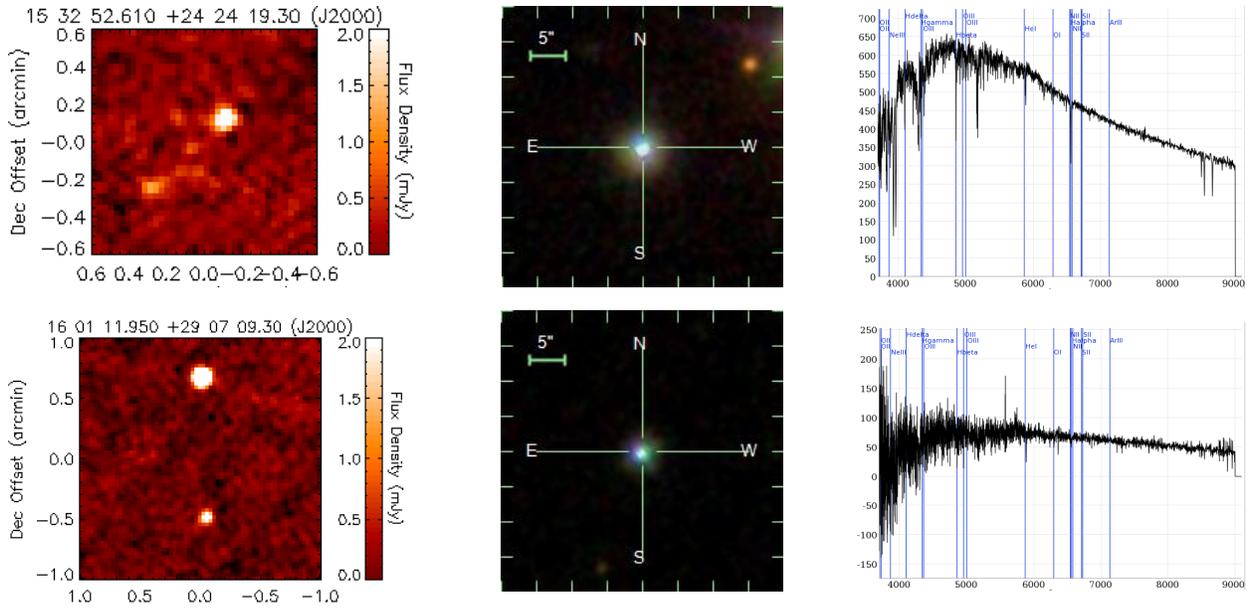

*Figure 2. Our best candidates for potential double-lobed radio stars. From top to bottom we show SDSS J023528.75+031651.0, J111403.64+415711.5, J153252.61+242419.3, and J160111.95+290703.3. Each row has the FIRST image on the left, SDSS in the middle, and the LAMOST spectrum on the right.*

**Misclassified Objects in LAMOST and Comparison with SDSS**

During our inspection of objects we found several wrongly classified spectra in LAMOST. Figure 3 shows three examples. The top row is the $z_{sp}$=1.127 QSO FBQS J104556.8+271759, the middle and bottom rows are the galaxies SDSS J113902.61+322820.7 and SDSS J151554.31+344357.3. SDSS spectra confirm their nonstellar nature.

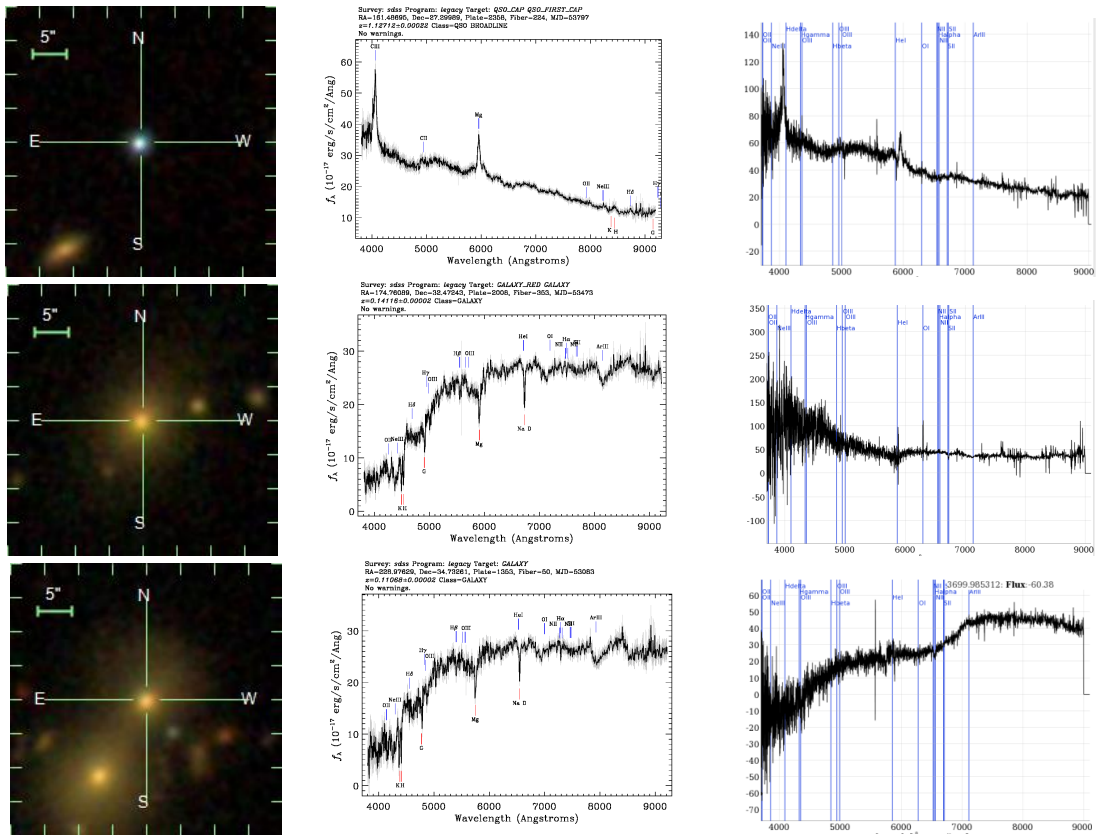

*Figure 3. Misclassified objects in LAMOST. Left: SDSS image, middle & right: SDSS & LAMOST spectra.*

This relatively high fraction of spectral type mismatches motivated us to crossmatch the 5,755,126 objects with spectra in LAMOST DR3 with the 4,851,200 objects with spectra in SDSS DR14, finding 46,627 matches within 3". Of these, 7875 (16.9 %) have LAMOST type *Star* and SDSS DR14 type *Galaxy* (7282) or *QSO* (593). Restricting the match radius to 1", the type mismatch remains at 16.8 % (7556/44,889), suggesting that the fraction of wrongly classified objects in the entire LAMOST survey may be similar. Of the 46,627 LAMOST-SDSS matches, 287 are classified as galaxies and 20 as QSOs in LAMOST, of which only 7 galaxies (2.2 %) are classified as stars in SDSS DR14: four are stars superimposed on galaxies and three are starforming galaxies with cz~500 km/s.

## CONCLUSIONS

In a cross-match of 2.5 million spectroscopic stars with almost one million radio sources from FIRST only very few candidates of stars with double radio lobes were found, confirming that such objects must be extremely rare. In fact, confirmation of the first real such association is still pending, and requires more sensitive radio images, deeper and higher-resolution optical images, possibly from space, as well as precise proper motions and parallaxes (e.g. from GAIA) for the candidate stars. A statistical assessment of the chance probability of these alignments is hampered by the fact that many spectroscopic targets were chosen exactly because of the presence of a radio source.

## ACKNOWLEDGEMENTS

We are grateful to M. A. Jimenez Valencia (Univ. de Sonora, Mexico) for sharing his FORTRAN code to find radio source straddling arbitrary sky positions, and to R. Coziol and V. Gámez R. for useful comments.

## BIBLIOGRAPHY

ABOLFATHI, B, AGUADO, D.S., AGUILAR, G., ALLENDE PRIETO, C., and 323 coauthors (2017), "The 14[th] Data Release of the Sloan Digital Sky Survey", https://arXiv.org/abs/1707.09322

ANDERNACH, H. (2015). "A new type of double-lobed radio-emitting stars?", oral pres. at 4th international TIGRE workshop. Guanajuato, see ftp://ftp.hs.uni-hamburg.de/pub/outgoing/uwolter/TIGRE-Slides_2015-16/2015-GTO

ANGUIANO, B. REBASSA-MANSERGAS, A. GARCIA-BERRO, E., et al. (2017). "The kinematics of the white dwarf population from the SDSS DR12", Mon. Not. Roy. astron. Soc., 469, 2102-2120

CUI X.-Q., ZHAO, Y.-H., CHU, Y.-Q., LI, G.-P. and 73 coauthors (2012). "The Large Sky Area Multi-Object Fiber Spectroscopic Telescope (LAMOST)", in Res. Astron. Astrophys., 12, 1197-1242

CUTRI, R.M., SKRUTSKIE, M.F., VAN DYK, S., BEICHMAN, C.A., et al. (2003). "2MASS All-Sky Catalog of Point Sources", UMass/IPAC (ftp://cdsarc.u-strasbg.fr/pub/cats/II/246/)

CUTRI, R.M., et al. (2013). "AllWISE Source Catalog", see http://wise2.ipac.caltech.edu/docs/release/allwise/expsup/sec2_1a.html

FLEWELLING, H.A., MAGNIER, E.A., CHAMBERS, K.C., et al. (2016). "The Pan-STARRS1 Database and Data Products", https://arxiv.org/abs/1612.05243

HELFAND, D.J., WHITE, R.L., BECKER, R.H., (2015). "The Last of FIRST: The Final Catalog and Source Identifications", Astrophys. J., 801, Art. 26

HJELLMING, R. M. (1988). "Radio Stars", in Galactic and Extragalactic Radio Astronomy, eds. G. L. Verschuur & K. I. Kellermann, Berlin: Springer, p. 381-438

JIMENEZ VALENCIA, M.A., ANDERNACH, H. (2015). "Búsqueda de estrellas coincidentes con radiofuentes dobles", XX Verano de la Investigación Científica y Tecnológica del Pacífico (Delfín)

KIMBALL, A.E., KNAPP, G.R., IVEZIĆ, Ž., WEST, A.A., BOCHANSKI, J.J., et al., (2009). "A Sample of Candidate Radio Stars in FIRST and SDSS", Astrophys. J., 701, 535-546

MACHALSKI, J., JAMROZY, M., ZOLA, S. (2001). "The new sample of Giant radio sources I.", Astron. Astrophys. 371, 445-469

MACHALSKI, J., JAMROZY, M., ZOLA, S., KOZIEL, D. (2006). "The new sample of Giant radio sources II.", Astron. Astrophys. 454, 85-94

MIRABEL, I.F, RODRÍGUEZ, L.F., (1999). Sources of Relativistic Jets in the Galaxy, Ann. Rev. Astron. & Astroph. 37, 409-443

ORTIZ MARTÍNEZ, A.F., ANDERNACH, H. (2016). "A Search for double-lobed radio emission from Galactic Stars and Spiral Galaxies", in Memoria del 18º Verano de la Ciencia, Región Centro (arXiv.org/abs/1610.02572)

TAYLOR, M., (2013). "TOPCAT: Tool for Operations on Catalogues and Tables", www.star.bris.ac.uk/~mbt/topcat

WENDKER, H. (2001). "Catalogue of Radio Stars", available at ftp://cdsarc.u-strasbg.fr/pub/cats/VIII/99

WHITE, R.L., BECKER, R.H., GREGG, M.D., et al. (2000). "The FIRST Bright Quasar Survey. II. 60 Nights and 1200 Spectra Later", in Astrophys. J. Suppl., 126, 133-207